\documentclass[journal=jacsat,manuscript=article]{achemso}

\usepackage[version=3]{mhchem}
\usepackage{titlesec}
\usepackage{float}
\usepackage[utf8]{inputenc}
\usepackage{graphicx}
\usepackage{amsmath}
\usepackage{mathtools}
\usepackage{geometry}
\usepackage{array}
\usepackage{csquotes}
\usepackage{soul}
\usepackage{braket}
\usepackage{xcolor}

\newcommand{\ev}[1]{\left\langle #1 \right\rangle}
\newcommand{\diota}{\dot{\iota}}

\author{Bilal Khalid}
\affiliation[Purdue University]{Department of Physics and Astronomy, Purdue University, West Lafayette, IN-47907, USA}
\author{Sabre Kais}
\email{skais@ncsu.edu}
\affiliation[North Carolina State University]{Department of Electrical and Computer Engineering, North Carolina State Univeristy, Raleigh, NC-27606, USA}

\title{A Classical Analogue of Entanglement for a Kicked Top}

\begin{document}

\begin{abstract}
Classical chaos is usually characterized by the sensitive dependence of trajectories on initial conditions. However, in quantum mechanics, the unitarity of time evolution ensures that the distances between quantum states are preserved in time. It has, therefore, been a challenge in quantum chaos to reconcile the lack of sensitive dependence on initial conditions with classical chaos. Ballentine has suggested a way out by constructing a parallel argument in classical mechanics based on the preservation of the inner product of phase space densities\cite{ballentine}. He has argued that the exponential separation of nearby states is a good identifier of chaos only at the level of individual trajectories. For statistical states such as quantum states and classical phase space densities, chaos must instead be identified by some fundamentally statistical signatures. The search for these signatures is the primary goal in quantum chaos research\cite{berry,gutzwiller,haake}. However, this perspective also naturally motivates the search for classical analogues of these signatures, to reveal the inner machinery of chaos in quantum systems. One widely recognized signature of chaos in quantum systems is the dynamical generation of entanglement. Chaos in the classical system is correlated with a greater entanglement production in the corresponding quantum system\cite{zurek1994, zurek1995, sarkar1998, furuya1998, sarkar1999-1, sarkar1999-2, pattanayak, paz, lakshminarayan2001-1, lakshminarayan2001-2, gong, lakshminarayan2004, dicke, furuya2004, giulio, bonanca, bsanders, ghose, chaudhury, google, ktm-eePRA}. One of the most well-studied examples of this is the kicked top model\cite{ktm-haake}. In this paper, we construct a classical analogue of bipartite entanglement in terms of the mutual information between phase space distributions of subsystems and find completely analogous signatures of chaos as those found in entanglement for the kicked top Hamiltonian.
\end{abstract}

\section{Introduction}
\label{introduction}

Quantum chaos is the study of the quantum mechanical properties characteristic of systems that exhibit chaos classically\cite{berry,gutzwiller,haake}. Traditionally, the primary focus in the field has been on the determination of universal features in the spectral statistics and the eigenstates of chaotic Hamiltonians. However, in recent years, developments in quantum information science and phenomenal advances in quantum simulation technologies have enabled novel theoretical and experimental avenues for exploring the dynamical manifestations of chaos in quantum systems. Information-theoretic measures such as entanglement entropy, quantum Fisher information, OTOCs (out-of-time-order correlators), etc. have been suggested as new probes for tracking quantum chaos. Consequently, a fresh understanding of quantum chaos has emerged that has revealed its fundamental significance in quantum dynamical processes, crucial to understanding decoherence, many-body systems and black hole physics, such as entanglement generation\cite{zurek1994, zurek1995, sarkar1998, furuya1998, sarkar1999-1, sarkar1999-2, pattanayak, paz, lakshminarayan2001-1, lakshminarayan2001-2, gong, lakshminarayan2004, dicke, furuya2004, giulio, bonanca, bsanders, ghose, chaudhury, google, ktm-eePRA}, information scrambling\cite{nima, maldacena, hosur, rozenbaum, monroe, xiao} and quantum thermalization\cite{deutsch, srednicki1994, rigol, science-thermalization, srednicki2019}.

The issue of quantum entanglement has been the subject of much debate since Einstein, Podolsky and Rosen pointed out the ``bizarre'' consequences it can lead to\cite{einstein}. Schr\"{o}dinger declared it as ``the characteristic trait of quantum mechanics, the one that enforces its entire departure from classical lines of thought\cite{schrodinger}.'' In its essence, entanglement expresses the \textit{nonlocal} and \textit{nonseparable} nature of quantum states in a form that is completely alien to classical physics\cite{myrvold}. In quantum information science, it has been identified as a central resource in quantum communication protocols, quantum cryptography and quantum information processing and storage\cite{nielsen-chuang, preskill-notes}. 

Remarkably, the dynamical generation of entanglement (within the system or with an environment) is intimately tied to the chaoticity properties of the underlying classical phase space. It has been observed that wave packets centered on regions of phase space that are classically chaotic yield a greater entanglement entropy production than classically regular regions. For chaotic initial conditions, the entanglement entropy grows linearly at a rate given by the sum of the positive Lyuapunov exponents, the classical Kolmogorov-Sinai entropy rate; whereas for the regular case, the entropy grows only logarithmically with time\cite{zurek1994, zurek1995, sarkar1998, furuya1998, sarkar1999-1, sarkar1999-2, pattanayak, paz, lakshminarayan2001-1, lakshminarayan2001-2, gong, lakshminarayan2004, dicke, furuya2004, giulio, bonanca, bsanders, ghose, chaudhury, google, ktm-eePRA}.

A system for which the chaos-entanglement relationship has been extensively studied is the kicked top model\cite{sarkar1999-2, lakshminarayan2004, bsanders, ghose, chaudhury, google, ktm-eePRA, ktm-haake}. In this system, the evolution of the angular momentum $\textbf{J}$ (``the top") is governed by two kinds of process: (i) precession of $\textbf{J}$ around a fixed axis at a constant rate and (ii) a periodic sequence of kicks that bring about an instantaneous change in $\textbf{J}$. The Hamiltonian for this system commutes with $\textbf{J}^2$, so the quantum evolution is confined within a subspace characterized by an eigenvalue $j(j+1)$ of $\textbf{J}^2$. Moreover, the model is chaotic in the classical limit $j \to \infty$. This model was introduced by Haake et al. to analyze how chaos arises as a system becomes more and more classical\cite{ktm-haake}.

A particularly interesting realization of this model is in terms of a collection of spins-$1/2$, where $\textbf{J}$ denotes the collective angular momentum of the spins. This approach has been used to study bipartite entanglement in the model as a function of time and initial state\cite{bsanders, ghose, chaudhury, google, ktm-eePRA}. In a common scenario, the system is initialized in a spin-coherent state i.e. a minimum uncertainty angular momentum state, and the growth of entanglement entropy of a single spin-$1/2$ is tracked. The growth of entropy has been found to carry strong signatures of chaos in the underlying classical dynamics: (i) for an initial state centered in a classically chaotic region of phase space, the entanglement entropy grows linearly at a rate given by the Lyuapunov exponent before reaching the saturation point, whereas, for the classically regular case, the entropy grows only logarithmically; (ii) for initial conditions centered in classically chaotic regions of phase space, the equilibrium entropy (also known as average entropy) is larger compared to those centered in classically regular regions\cite{bsanders, ghose, chaudhury, google, ktm-eePRA}. 

With recent advances in a variety of quantum simulation platforms, there has also been a lot of interest in experimental investigations of this correlation. A quantum simulation of the kicked top was achieved by Chaudhury et al. using the $F=3$ hyperfine ground state of $^{133} \text{Cs}$\cite{chaudhury}. In their experiment, the total angular momentum in the Hamiltonian was taken to be the sum of the electron and nuclear spins of a single $^{133} \text{Cs}$ atom. Consequently, the theoretically predicted correspondence between entanglement, as quantified by the linear entropy of the electron spin, and classical chaos was corroborated. Later, similar conclusions were obtained by Neill et al. in their quantum simulation experiment of the same Hamiltonian using a three-qubit ring of planar transmons\cite{google}. Commenting on their findings, they added, ``it is interesting to note that chaos and entanglement are each exclusive to their respective classical and quantum domains, and any connection is surprising.''

The connection is surprising because a purely quantum property (entanglement) is being related with a purely classical one (chaos), each one understood to have no counterpart on the other side. The standard argument for the absence of chaos in quantum mechanics proceeds like this. Suppose $\ket{\psi_1 (0)}$ and $\ket{\psi_2 (0)}$ represent two initially close quantum states i.e. $\braket{\psi_1 (0) | \psi_2 (0)} = 1 - \epsilon$ ($\epsilon$ being a small number.) Under unitary evolution of $\ket{\psi_1 (0)}$ and $\ket{\psi_2 (0)}$, we should have $\braket{\psi_1 (t) | \psi_2 (t)} = 1 - \epsilon$ for all times $t$. So, the states do not separate in time and this is taken to imply that there can be no chaos in quantum mechanics\cite{ballentine}. However, Ballentine has argued that a parallel argument can be constructed in classical mechanics too if classical states are taken to be represented by probability distributions in phase space. For two phase space distributions $\rho_1(q,p,t)$ and $\rho_2(q,p,t)$, the construction $\{\rho_1(t)|\rho_2(t)\} = \int \int \rho_1(q,p,t) \rho_2(q,p,t) \: dq dp$ is a well-defined inner product on phase space and is invariant under the Liouvillian dynamics of $\rho_1$ and $\rho_2$. But no one can deny the existence of chaos in classical mechanics. Ballentine then concludes that the confusion about quantum chaos is merely a reflection of the confusion about the notion of ``state'' in classical and quantum mechanics. The more adequate classical analogue of a quantum state is not a single trajectory but a phase space distribution, and chaos in such states must be identified by some statistical signatures\cite{ballentine}.

One such signature is the growth of entanglement in quantum systems as discussed above. This naturally raises the question of what would be a good classical analogue of entanglement in the statistical interpretation of classical physics. Constructing such an analogue is desirable for two related reasons: (i) a comparison between conceptually similar identifiers of chaos across the classical-quantum divide can enable a fresh understanding of the classical-quantum correspondence, especially in light of the issues raised by chaos; (ii) since quantum chaos is still far from understood, an analysis of a classical analogue of a quantum signature of chaos can reveal the inner machinery of quantum chaos, that would otherwise be hidden from view.

To construct this analogue, it would be convenient to consider the meaning of entanglement in the Wigner function formalism of quantum mechanics as it provides a visualization of quantum states in phase space. In this formulation, the state of a quantum system is represented by a real-valued function in phase space $W(q,p)$, called the Wigner function. This function in many ways acts like the classical phase space density $\rho(q,p)$. However, an important difference is that $W(q,p)$ is not really a distribution as it can take negative values unlike $\rho(q,p)$\cite{qformulations,wignerfunc}.

In the Wigner function formalism, two systems are entangled iff their collective Wigner function is nonseparable i.e. if $W(q_1,q_2,p_1,p_2)$ is the Wigner function of the total system and $W_1(q_1,p_1) = \int \int W(q_1,q_2,p_1,p_2) \: d q_2 d p_2$ and $W_2(q_2,p_2) = \int \int W(q_1,q_2,p_1,p_2) \: d q_1 d p_1$ are the Wigner functions of systems $1$ and $2$ respectively, then $W(q_1,q_2,p_1,p_2) \neq W_1(q_1,p_1) \times W_2(q_2,p_2)$. This motivates the construction of a classical analogue in terms of the separability of phase space density $\rho$. The classical state is separable iff $\rho(q_1,q_2,p_1,p_2) = \rho_1(q_1,p_1) \times \rho_2(q_2,p_2)$ and is nonseparable otherwise, where $\rho_1(q_1,p_1) = \int \rho(q_1,q_2,p_1,p_2) \: d q_2 d p_2$ and $\rho_2(q_2,p_2) = \int \rho(q_1,q_2,p_1,p_2) \: d q_1 d p_1$. To quantify the degree of nonseparability, we will use mutual information which for two random variables $X_1$ and $X_2$ is defined as $I_{12} = \int \int \rho(X_1,X_2) \: \text{log} \Big[ {\rho(X_1,X_2)}/\big({\rho_1 (X_1) \rho_2 (X_2)\big)} \Big] dX_1 dX_2$. 

The mutual information $I_{12}$ between two random variables $X_1$ and $X_2$ is a non-negative number i.e. $I_{12} \geq 0$. The measure $I_{12}=0$ if and only if the joint probability distribution of $X_1$ and $X_2$ is completely separable i.e. $\rho(X_1,X_2) = \rho_1 (X_1) \rho_2 (X_2)$. Moreover, $I_{12}$ becomes infinite when the two variables are perfectly correlated i.e. if there exists a functional relationship between $X_1$ and $X_2$. While mutual information is not the only possible measure of classical nonseparability, we have employed it for convenience of computation. For other possible measures, see refs.~\cite{sarkar1998, lakshminarayan2001-1, furuya2004, giulio, eberly}.

In this paper, we have analyzed the growth of mutual information in the classical kicked top. We bipartition the total angular momentum $\textbf{J}$ into two parts $\textbf{J}_1$ and $\textbf{J}_2$ and compute the mutual information between the variables on the two sides of the partition. We find striking resemblances between the growth of mutual information and the bipartite entanglement. Mutual information, like entanglement, carries clear signatures of chaos in the underlying dynamics. Under chaotic dynamics, it grows linearly at a rate proportional to the Lyuapunov exponent. Whereas, for regular dynamics, the growth starts to slow down well before equilibrium is attained. Similarly, initial states centered in chaotic regions of phase space end up with a higher mutual information at equilibrium compared to regular regions, in complete analogy with bipartite entanglement.

The organization of the paper is as follows. In Sec.~\ref{classicaldynamics}, we introduce the kicked top Hamiltonian and describe its classical dynamics. In Sec.~\ref{quantumentanglement}, we recall the correspondence between entanglement and classical dynamics. Then, in Sec.~\ref{secmutualinfo}, we present our calculations for classical mutual information. Finally, in Sec.~\ref{summary}, we provide a summary of the results and an outlook for the future.

\section{Classical Dynamics}
\label{classicaldynamics}

Consider the angular momentum operator $\hbar \textbf{J} = \hbar (J_x, J_y, J_z)$ satisfying the commutation relations $[J_i,J_j] = \diota \varepsilon_{ijk} J_k$. The Hamiltonian for the kicked top is then expressed in terms of $\textbf{J}$ as\cite{ktm-haake},
\begin{equation}
    H(t) = \frac{\hbar p}{\tau} J_y + \frac{\hbar \kappa}{2 j} J_z^2 \sum_{n=- \infty}^{+ \infty} \delta(t - n \tau).
\label{hamiltonian}
\end{equation}
The first term describes the precession of the rotor around the y-axis at a rate $p/\tau$. The second term represents a periodic sequence of kicks separated by a period $\tau$. Intuitively, this term can be thought of as a sudden precession around the z-axis by an angle proportional to $J_z/j$, where $j$ is the total angular momentum quantum number. Once we initialize our system in the subspace characterized by the eigenvalue $j(j+1)$ of the operator $\textbf{J}^2$, we stay within the same subspace for all times since $[\textbf{J}^2,H(t)] = 0$. $\kappa$ is a dimensionless constant which controls the strength of the kick. For this paper, we are going to choose $p=\pi/2$ i.e. the top precesses around the y-axis by an angle $\pi/2$ between successive kicks.

\begin{figure}[t]
\centering
\includegraphics[width=\linewidth]{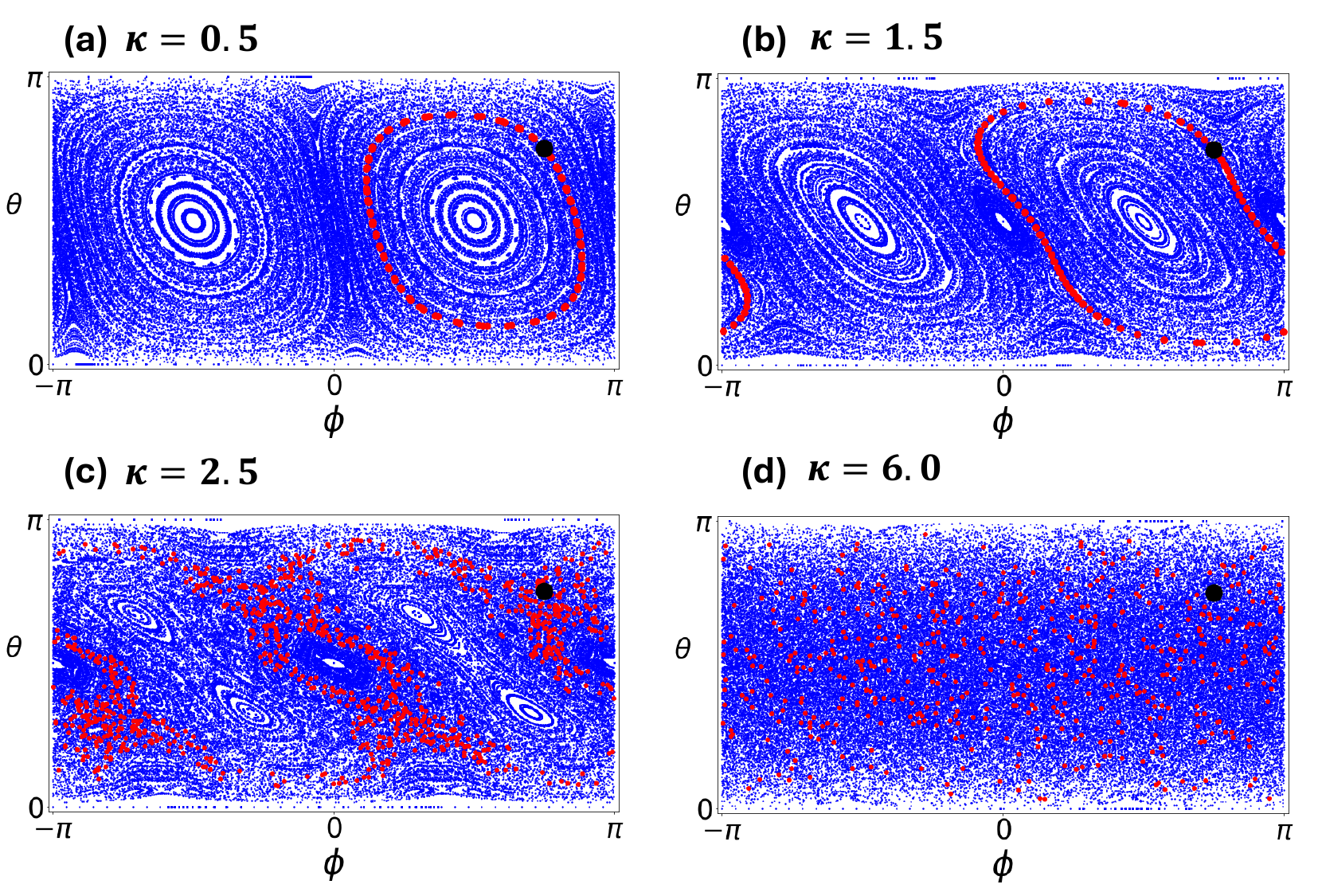}
\caption{\textbf{Classical phase portraits for the kicked top.} The trajectories of rescaled angular momenta $\textbf{X} = \textbf{J}/j$ in the classical limit $j \to \infty$, represented in terms of the polar and the azimuthal angles on a unit sphere. As $\kappa$ is tuned from low to high, an order-to-chaos transition occurs in the phase space. Red markers represent the trajectories corresponding to the initial condition $\theta_0 = 3\pi/4, \phi_0 = 3\pi/4$ (black marker.)}
\label{classicalphasespace}
\end{figure}

Working in the Heisenberg picture, we are interested in tracking the evolution of $\textbf{J}$ in time. The evolution of the operator $J_i$ in $n$ time steps can be represented as $J_i^{(n)} = (U^\dagger) ^ n \: J_i \: U^n$, where $U$ is the unitary evolution corresponding to the interval $\tau$ between successive kicks\cite{ktm-haake},
\begin{equation}
    U = e^{-\diota (\kappa/2 j) J_z^2} e^{-\diota (\pi/2) J_y}.
\label{unitaryevolution}
\end{equation}
The evolution of $\textbf{J}$ can be represented in terms of the following non-linear operator recursion relations which determine how $\textbf{J} = \textbf{J} ^ {(i)}$ is updated to $\textbf{J}' = \textbf{J} ^ {(i+1)}$ after each time step\cite{ktm-haake},
\begin{align}
    J_{x}' &= \frac{1}{2} (J_z + \diota J_y) \: \text{e}^{- \diota \frac{\kappa}{j} (J_x - \frac{1}{2})} + \text{h.c.} \nonumber \\
    J_{y}' &= \frac{1}{2 \diota} (J_z + \diota J_y) \: \text{e}^{- \diota \frac{\kappa}{j} (J_x - \frac{1}{2})} + \text{h.c.} \label{qrecur} \\
    J_{z}' &= - J_x \nonumber.
\end{align}
Defining the rescaled angular momentum as $\textbf{X}=\textbf{J}/j$ and taking the classical limit $j \to \infty$, we can track the evolution of the now real-valued $\textbf{X}=(X,Y,Z)$ on the surface of a unit sphere using the following recursion relations obtained from \eqref{qrecur},
\begin{align}
    X' &= \text{Re} \{ (Z + \diota Y) \: e^{- \diota \kappa X} \} \nonumber \\
    Y' &= \text{Im} \{ (Z + \diota Y) \: e^{- \diota \kappa X} \} \label{crecur} \\
    Z' &= - X \nonumber.
\end{align}

In Fig.~\ref{classicalphasespace}, we have plotted some examples of the phase portraits in spherical coordinates (i.e. $X=\sin \theta \cos\phi$, $Y=\sin \theta \sin\phi$ and $Z=\cos\theta$) that are produced by the recursion relations \eqref{crecur} for different values of the kick strength $\kappa$. As we increase the kick strength $\kappa$, chaos emerges in the phase space and islands of regularity begin to shrink. Eventually, for a large enough value of $\kappa$, chaos completely takes over.

\section{Quantum Entanglement}
\label{quantumentanglement}

\begin{figure}[t]
\centering
\includegraphics[width=0.6\linewidth]{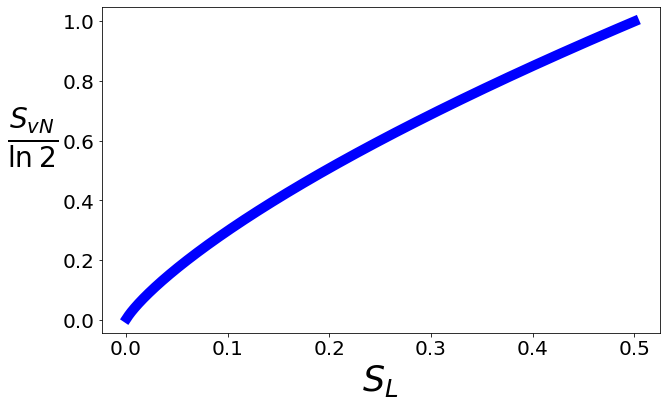}
\caption{\textbf{von Neumann entropy and linear entropy for spin-$1/2$ systems.}} 
\label{vNvsL}
\end{figure}

Consider a collection of $N$ spins-$1/2$ with the corresponding spin operators $\textbf{S}_i=(S_{ix},S_{iy},S_{iz})$ such that the dynamics of the total angular momentum $\textbf{J} = \sum_{i=1}^{N} \textbf{S}_i$ is governed by the Hamiltonian \eqref{hamiltonian}. In terms of the spin operators $\textbf{S}_i$, the Hamiltonian can be re-written as
\begin{equation}
    H(t) = \frac{\hbar \pi}{2 \tau} \sum_{i=1}^{N} S_{iy} + \frac{\hbar \kappa}{2 j} \Bigg(\sum_{i=1}^{N} S_{iz}^2 + \sum_{i \neq j} S_{iz} S_{jz} \Bigg) \sum_{n=- \infty}^{+ \infty} \delta(t - n \tau).
\label{spinhamiltonian}
\end{equation}
Before each kick, each spin independently precesses around the y-axis by an angle $\pi/2$. Noting that $(\sum_{i=1}^{N} S_{iz}^2 + \sum_{i \neq j} S_{iz} S_{jz}) = J_z \: (\sum_{i=1}^{N} {S}_{iz})$, the kick can be understood as causing a sudden precession of each spin around the z-axis by an angle proportional to $J_z/j$, a collective variable of the system.

\begin{figure}[h!]
\centering
\includegraphics[width=\linewidth]{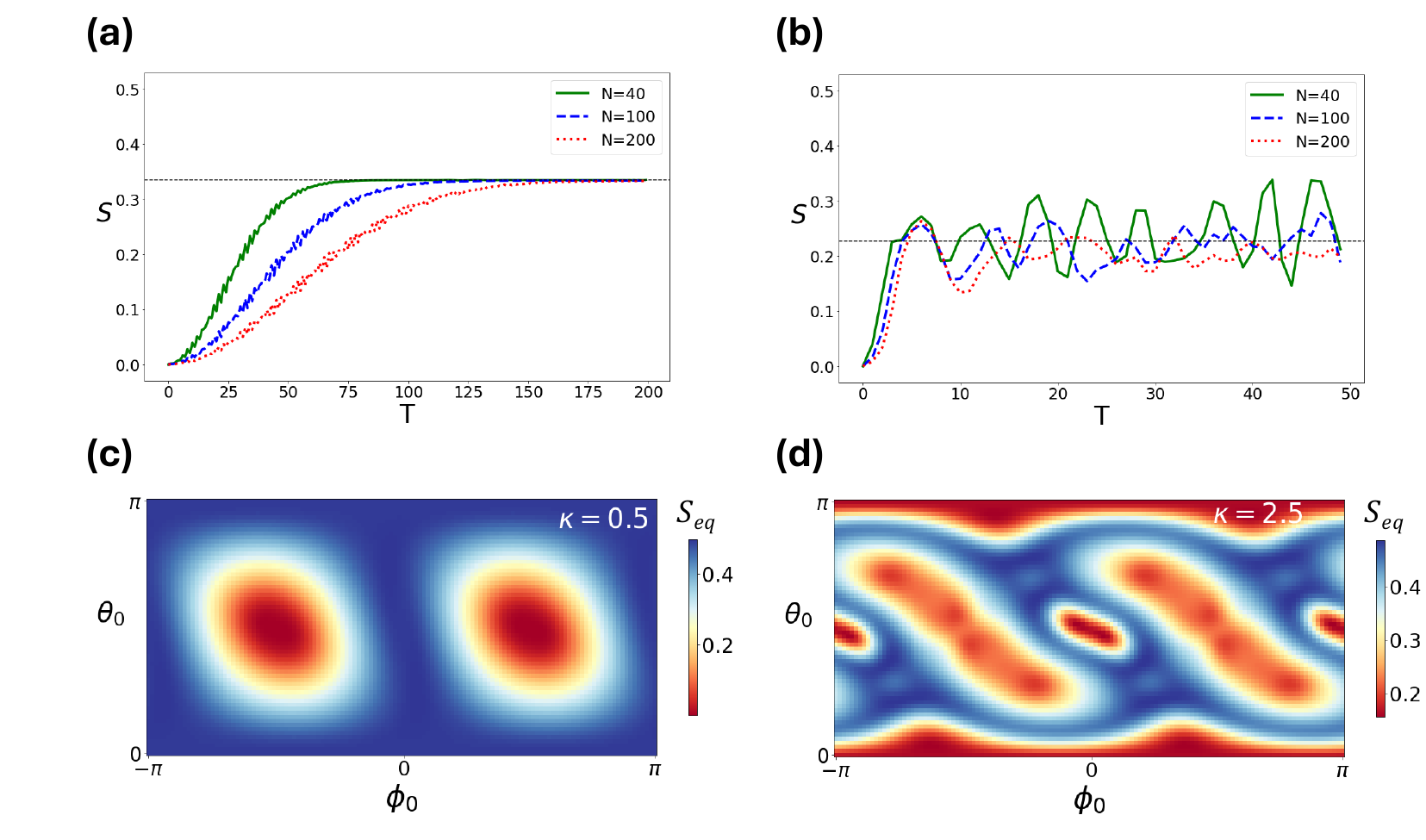}
\caption{\textbf{Linear entropy.} Linear entropy  of a single spin $S = 1 - \text{Tr}_1 (\rho_1^2)$ as a function of time steps $T$ and initial orientation $(\theta_0,\phi_0)$. (a) and (b) show the time dynamics of $S$ with the initial orientation $(\theta_0=3 \pi/4, \phi_0=3 \pi/4)$ for $\kappa=0.5$ and $\kappa=2.5$ respectively. (c) and (d) display the equilibrium value $S_{eq}$ of linear entropy as a function of the initial orientation $(\theta_0,\phi_0)$ for $\kappa=0.5$ and $\kappa=2.5$ respectively. The system size is taken to be $N=40$. $S_{eq}$ is estimated by averaging $S$ over an appropriate time interval after reaching saturation. For (c), the average is performed for $60 \leq T \leq 100$ whereas for (d), the average is computed over $20 \leq T \leq 40$. There is a striking resemblance of the plots (c) and (d) with the corresponding classical phase portraits shown in Figs.~\ref{classicalphasespace}(a) and (c).} 
\label{linearentropy}
\end{figure}

In this section, we recall the dynamics of bipartite entanglement generated by this Hamiltonian. We initialize the system in the spin-coherent state,
\begin{equation}
    \ket{\psi(t=0)} = \bigotimes_{i=1}^{N} \ket{\theta_0,\phi_0}_i = \text{exp} \{\diota \theta_0 (J_x \: \text{sin} \: \phi_0 - J_y \: \text{cos} \: \phi_0) \} \ket{j,j}.
\label{inistatej}
\end{equation}
This is the minimum uncertainty angular momentum state pointing along a certain direction $(\theta_0,\phi_0)$ for a given total angular momentum quantum number $j$. For $N$ spins-$1/2$ pointing in the same direction, we have $j=N/2$. $\ket{\theta_0,\phi_0}$ is the spin-$1/2$ state pointing along $(\theta_0,\phi_0)$ on the Bloch sphere i.e. $\ket{\theta_0,\phi_0} = \text{cos} (\theta_0/2) \ket{\uparrow} + e^{-\diota \phi_0} \text{sin}(\theta_0/2) \ket{\downarrow}$. The initial state is completely separable, however, entanglement is generated as a result of the unitary evolution \eqref{unitaryevolution}.

To track the dynamics of bipartite entanglement, we use linear entropy of a single spin-$1/2$ defined as $S = 1 - \text{Tr}_1 (\rho_1^2)$, where $\rho_1$ is the reduced density matrix for a single spin. $S=0$ for a pure state, and is maximized at $S=0.5$ for a completely mixed state. This measure is used only for convenience; qualitatively, the results are expected to be independent of the choice for pure states\cite{bsanders}. Even quantitatively, there is a nearly linear relationship between von Neumann entropy and linear entropy for spin-$1/2$ states as shown in Fig.~\ref{vNvsL}.

In Figs.~\ref{linearentropy}(a) and (b), we have plotted the time dynamics of entropy for the regular ($\kappa=0.5$) and chaotic ($\kappa=2.5$) scenarios, respectively. For both cases, the initial state is centered at $(\theta_0=3 \pi/4, \phi_0=3 \pi/4)$. The dynamics has been plotted for three different system sizes $N = 40,\: 100,\: 200$. For both scenarios, entropy grows consistently before saturating after some time $T_{eq}$. For the regular case, $T_{eq}$ increases with the size of the system $N$ as $O(\sqrt{N})$. On the other hand, the increase is only logarithmic $O(\ln N)$ for the chaotic case\cite{ktm-eePRA}. Moreover, the rate of entropy growth in the regular case starts to slow down well before reaching saturation, signifying a logarithmic growth of entropy. However, for the chaotic case, the growth is linear at a rate given by the Lyuapunov exponent\cite{ktm-eePRA}. For larger times, the entropy undergoes sequences of collapses and revivals, which recede into the indefinite future as the system size approaches the thermodynamic limit.

\begin{figure}[t]
\centering
\includegraphics[width=\linewidth]{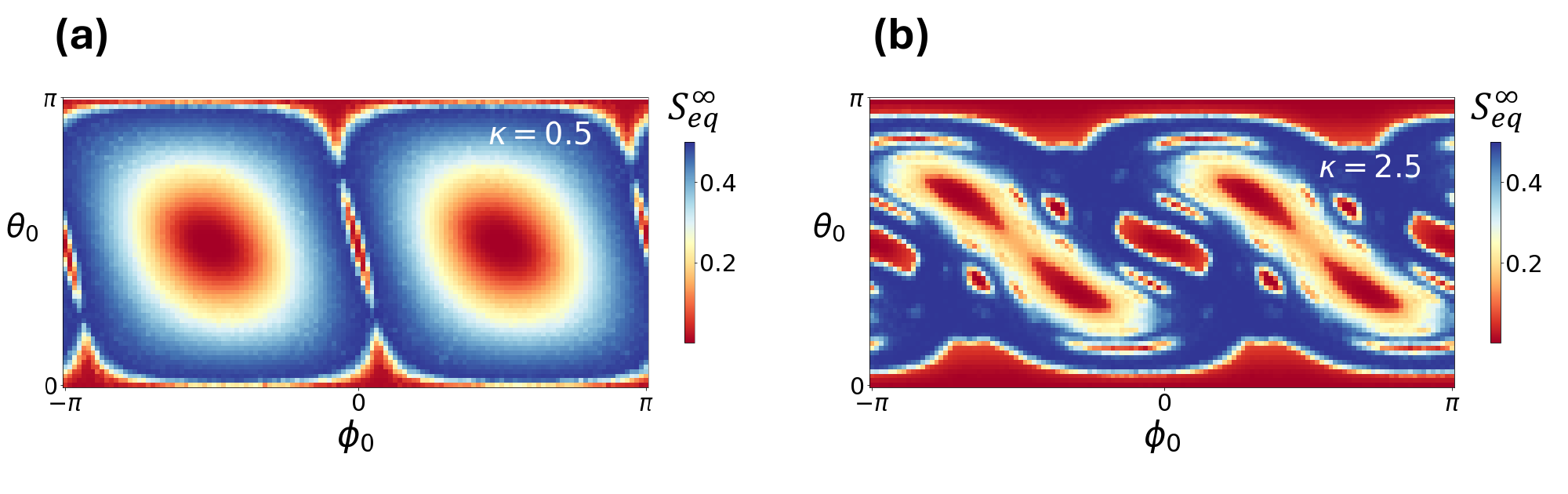}
\caption{\textbf{Linear entropy in the thermodynamic limit.} We have estimated the equilibrium value of linear entropy in the thermodynamic limit $S_{eq}^{\infty}$ as a function of the initial orientation $(\theta_0,\phi_0)$. For each $(\theta_0,\phi_0)$, $S^{\infty}=\ev{(\Delta X)^2}/2$ is computed classically by evolving $200$ trajectories sampled from a region of angular spread $\sin \theta_0 \Delta \theta \Delta \phi=1/j$ for $j=100$ centered at $(\theta_0,\phi_0)$. For both plots, $S^{\infty}$ is averaged between $400 \leq T \leq 500$.} 
\label{thermlimit}
\end{figure}

In Figs.~\ref{linearentropy}(c) and (d), equilibrium values of entropy $S_{eq}$ have been plotted as a function of the initial orientations $(\theta_0,\phi_0)$ for $N=40$. $S_{eq}$ is estimated by averaging the entropy over a chosen time interval after saturation. Remarkably, the plots of entanglement reflect the structure of the classical phase space in Figs.~\ref{classicalphasespace}(a) and (c). For $\kappa=2.5$ specifically, we find that the regions of chaos in classical phase space correspond to regions of higher entropy on the quantum side and the regions of regularity correspond to a lower entropy\cite{google}.

Finally, to obtain an estimate of entropy in the thermodynamic limit, we note that linear entropy for a state symmetric with respect to all the spins can be expressed as\cite{ghose},
\begin{equation}
    S = 1 - \text{Tr}_1 (\rho_1^2) = \frac{1}{2} \Big[1 - \frac{1}{j^2} (\ev{J_x}^2 + \ev{J_y}^2 + \ev{J_z}^2) \Big].
\label{Sestimationclassical}
\end{equation}
As $j \to \infty$, this becomes $S = \ev{(\Delta X)^2}/2$ where $\ev{(\Delta X)^2} = (\ev{\textbf{J}^2} - \ev{\bf{J}}^2) / j^2$. We can then compute $\ev{(\Delta X)^2}$ in the classical limit to estimate $S$ in the thermodynamic limit. The results for this calculation are shown in Fig.~\ref{thermlimit}. For each $(\theta_0,\phi_0)$, we evolved $200$ trajectories initialized in a region of angular spread $\sin \theta_0 \Delta \theta \Delta \phi=1/j$ centered at $(\theta_0,\phi_0)$ to calculate $\ev{(\Delta X)^2}$. These plots contain some extra minima regions (i.e. red regions) located around the fixed points of the classical phase space [Figs.~\ref{classicalphasespace}(a) and (c)] that were not captured in Figs.~\ref{linearentropy}(c) and (d).

\section{Classical Mutual Information}
\label{secmutualinfo}
In the \textit{Introduction}, we have motivated a classical notion of nonseparability quantified by mutual information. In this section, we use that measure to track nonseparability in the classical kicked top. Suppose we bipartition the system by dividing $\textbf{J}$ into $\textbf{J}_1$ and $\textbf{J}_2$ so that $\textbf{J}=$\textbf{J}$_1+$\textbf{J}$_2$. The Hamiltonian \eqref{hamiltonian} can then be re-expressed in terms of $\textbf{J}_1$ and $\textbf{J}_2$ as
\begin{equation}
    H(t) = \frac{\hbar \pi}{2 \tau} (J_{1y} + J_{2y}) + \frac{\hbar \kappa}{2 j} (J_{1z}^2+J_{2z}^2+2 J_{1z} J_{2z}) \sum_{n=- \infty}^{+ \infty} \delta(t - n \tau).
\end{equation}
$\textbf{J}_1^2$ and $\textbf{J}_2^2$ are conserved quantities since $[\textbf{J}_{1,2}^2,H(t)] = 0$. The unitary evolution operator over one cycle is $U=U_{z^2} U_{12} U_{y}$ where $U_{z^2} = e^{-\diota (\kappa/2 j) J_{1z}^2} e^{-\diota (\kappa/2 j) J_{2z}^2}$, $U_{12} = e^{-\diota (\kappa/j) J_{1z} J_{2z}}$ and $U_y = e^{-\diota (\pi/2) J_{1y}} e^{-\diota (\pi/2) J_{2y}}$. We can compute $\textbf{J}'_1 = U^\dagger \textbf{J}_1 U$ to produce the following recursion relations for the update of angular momentum of subsystem $1$ (see supporting information \ref{recursion},)
\begin{align}
    J'_{1x} &= \frac{1}{2} (J_{1z} + \diota J_{1y}) \: \text{e}^{- \diota \frac{\kappa}{j} (J_{1x} + J_{2x} + \frac{1}{2})} + \text{h.c.} \nonumber \\
    J'_{1x} &= \frac{1}{2 \diota} (J_{1z} + \diota J_{1y}) \: \text{e}^{- \diota \frac{\kappa}{j} (J_{1x} + J_{2x} + \frac{1}{2})} + \text{h.c.} \\
    J'_{1x} &=  - J_{1x} \nonumber.
\end{align}
\begin{figure}[t]
\centering
\includegraphics[width=\linewidth]{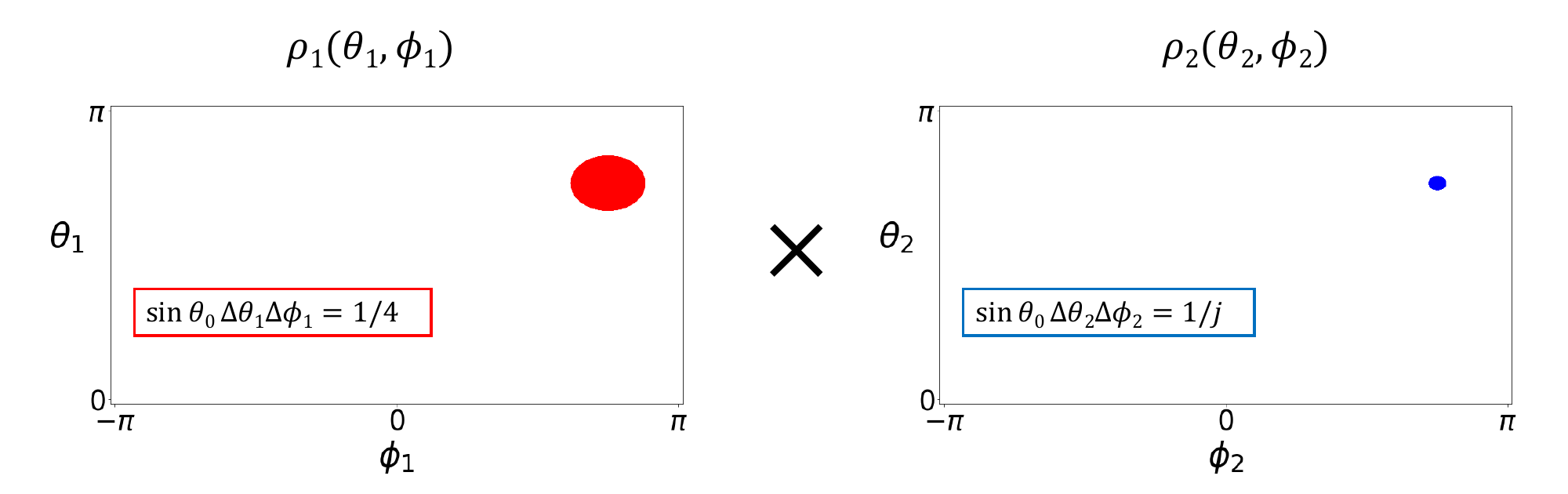}
\caption{\textbf{Initial distribution for mutual information calculations.} The initial distribution for the total system $\rho_{12}(\theta_1,\phi_1,\theta_2,\phi_2) = \rho_1(\theta_1,\phi_1) \times \rho_2(\theta_2,\phi_2)$ for $(\theta_0=3 \pi/4, \phi_0=3 \pi/4)$ where $j=100$. Both $\rho_1$ and $\rho_2$ are uniformly distributed in the corresponding regions shaded in red and blue respectively. The distribution $\rho_{12}$ is chosen in analogy with the quantum state in \eqref{inistatej}. As time dynamics are generated using eqs.~\eqref{cprecur}, the distribution $\rho_{12} (t)$ is generally no longer separable i.e. $\rho_{12}(\theta_1,\phi_1,\theta_2,\phi_2,t) \neq \rho_1(\theta_1,\phi_1,t) \times \rho_2(\theta_2,\phi_2,t)$.} 
\label{inidist}
\end{figure}
For $\textbf{J}'_2$, we only need to interchange the indices $1$ and $2$ in the above equations. Finally, defining $\textbf{X}_{1,2}=\textbf{J}_{1,2}/j$ and taking the classical limit $j \to \infty$ we get
\begin{align}
    X'_1 &= \text{Re} \{ (Z_1 + \diota Y_1) \: e^{- \diota \kappa (X_1 + X_2)} \} \nonumber \\
    Y'_1 &= \text{Im} \{ (Z_1 + \diota Y_1) \: e^{- \diota \kappa (X_1 + X_2)} \} \label{cprecur} \\
    Z'_1 &= - X_1 \nonumber.
\end{align}
\begin{figure}[t]
\centering
\includegraphics[width=\linewidth]{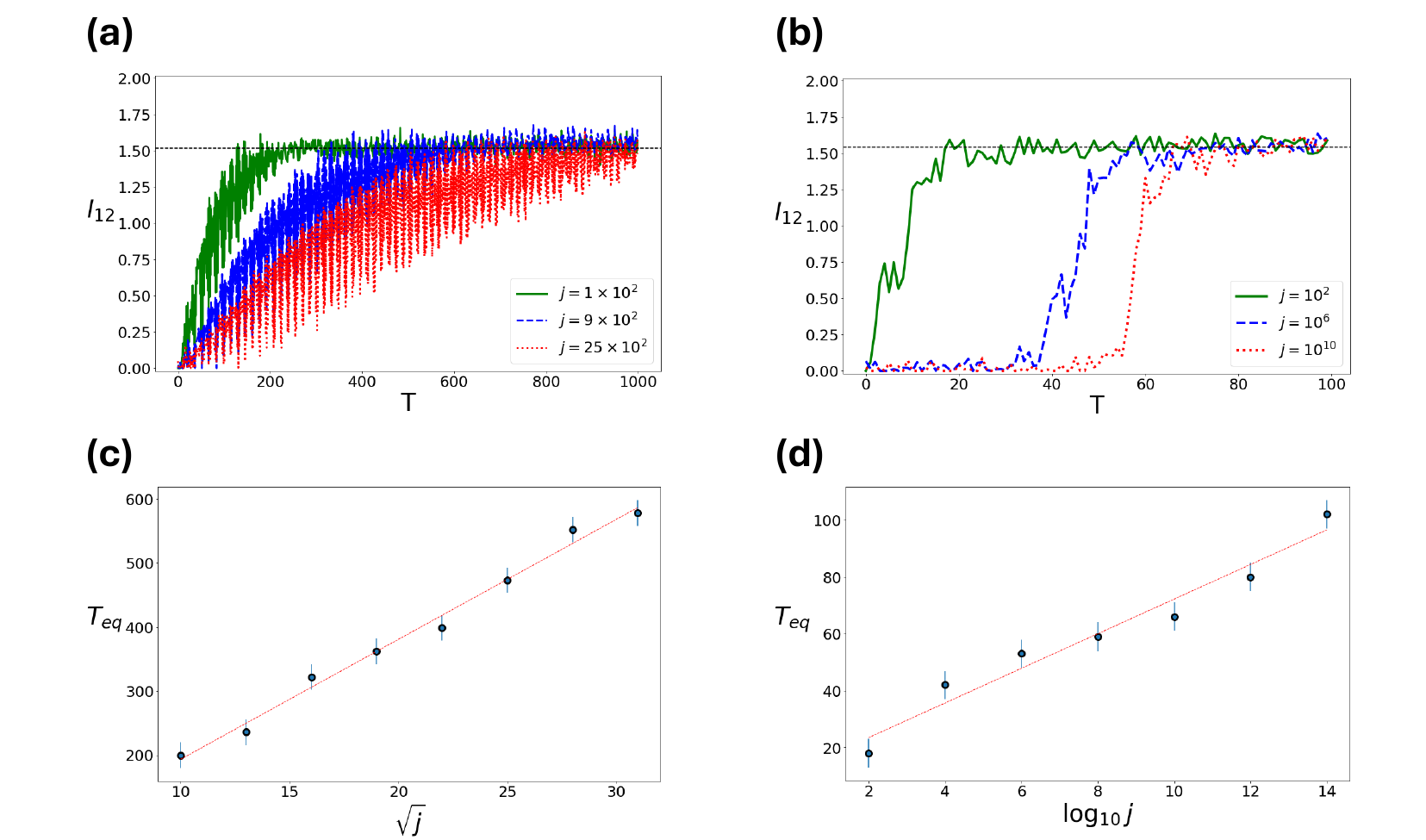}
\caption{\textbf{Mutual information growth and system size.} Mutual information $I_{12}$ between the variables $X_1=J_{1x}/j$ and $X_2=J_{2x}/j$ with initial orientation $(\theta_0=3 \pi/4,\phi_0=3 \pi/4)$ for $\kappa=0.5$ [(a) and (c)] and $\kappa=2.5$ [(b) and (d)], respectively. The system starts in a completely separable distribution with angular spread $\sin \theta_0 \Delta \theta \Delta \phi = 1/4$ for subsystem $1$ and $\sin \theta_0 \Delta \theta \Delta \phi = 1/j$ for subsystem $2$. A sample of $500$ points is drawn from this distribution and the corresponding trajectories are evolved to compute the statistics. (a) and (b) show the growth of $I_{12}$ with time, whereas, (c) and (d) display the advancement in equilibration time $T_{eq}$ with system size $j$.} 
\label{miwithsize}
\end{figure}
We take subsystem $1$ to be the analogue of a spin-$1/2$, while subsystem $2$ represents the rest of the system. This motivates our choice $\| \textbf{J}_1 \|=1/2$ and $\| \textbf{J}_2 \| = j - \| \textbf{J}_1 \|$. We initialize the system in a completely separable distribution i.e. the distribution for the total system is simply a product of the marginal distributions for subsystems $1$ and $2$. Both marginal distributions are taken to be uniformly distributed around $(\theta_0,\phi_0)$. For subsystem $2$, the angular spread of the initial distribution is taken to be $\sin \theta_0 \Delta \theta \Delta \phi = 1/j$, in analogy with the quantum state \eqref{inistatej}. On the other hand, for subsystem $1$, the initial angular spread is fixed at $\sin \theta_0 \Delta \theta \Delta \phi = 1/4$ (see Fig.~\ref{inidist}.) We sample initial conditions from this initial distribution for the total system, evolve them into trajectories and estimate the mutual information $I_{12}$ between the variables $X_1=J_{1x}/j$ and $X_2=J_{2x}/j$ based on $k$-nearest neighbor statistics\cite{mi_alg}.

\begin{figure}[t]
\centering
\includegraphics[width=\linewidth]{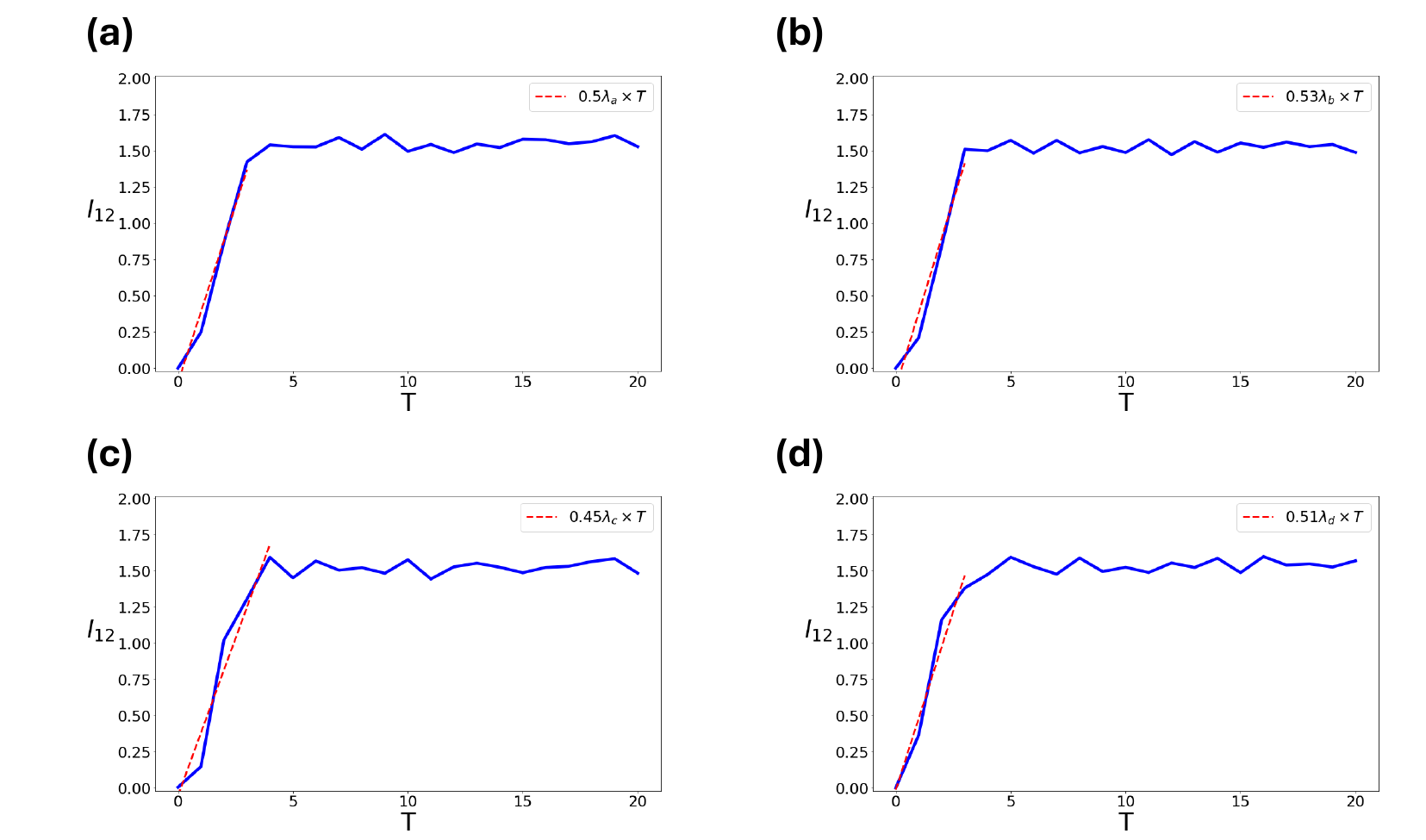}
\caption{\textbf{Mutual information growth and Lyuapunov exponents for a fully chaotic phase space.} A comparison of the growth rate of $I_{12}$ with the corresponding Lyuapunov exponents for four different cases at $\kappa=6.0$: (a) $(\theta_0=3 \pi/4, \phi_0=3 \pi/4)$, $\lambda_a = 0.978$; (b) $(\theta_0=\pi/3, \phi_0=2 \pi/3)$, $\lambda_b = 0.976$; (c) $(\theta_0=1.0, \phi_0=\pi/10)$, $\lambda_c = 0.974$; (d) $(\theta_0=\pi/4, \phi_0=\pi/3)$, $\lambda_d = 0.976$. For all these scenarios, $j=100$, and $1000$ samples are drawn from the initial distribution.} 
\label{miwithtime6.0}
\end{figure}

In Fig.~\ref{miwithsize}, we have shown the dynamics of $I_{12}$ for different system sizes $j$ (recall $j=N/2$.) Fig.~\ref{miwithsize}(a) shows the growth of $I_{12}$ for regular classical dynamics. The rate of growth decays with time, a signature of logarithmic growth. Moreover, as the system size $j$ increases, the growth slows further as the system is expected to take longer to reach equilibrium. Plot (c) shows that the equilibration time $T_{eq}$ increases as $O(\sqrt{j})$ with $j$. All these trends are completely analogous to the growth of quantum entropy for classically regular phase space.

An interesting feature of Fig.~\ref{miwithsize}(a) is that $I_{12}$ undergoes rapid oscillations toward its growth to saturation. These oscillations do not result from any estimation inaccuracies; rather, they seem to be a characteristic feature of dynamics for initial conditions located in classically regular regions of phase space. These oscillations are mirrored on the quantum side as well in the growth of linear entropy as system sizes become large (see supporting information \ref{oscilaltionsmi}.) 

\begin{figure}[t]
\centering
\includegraphics[width=\linewidth]{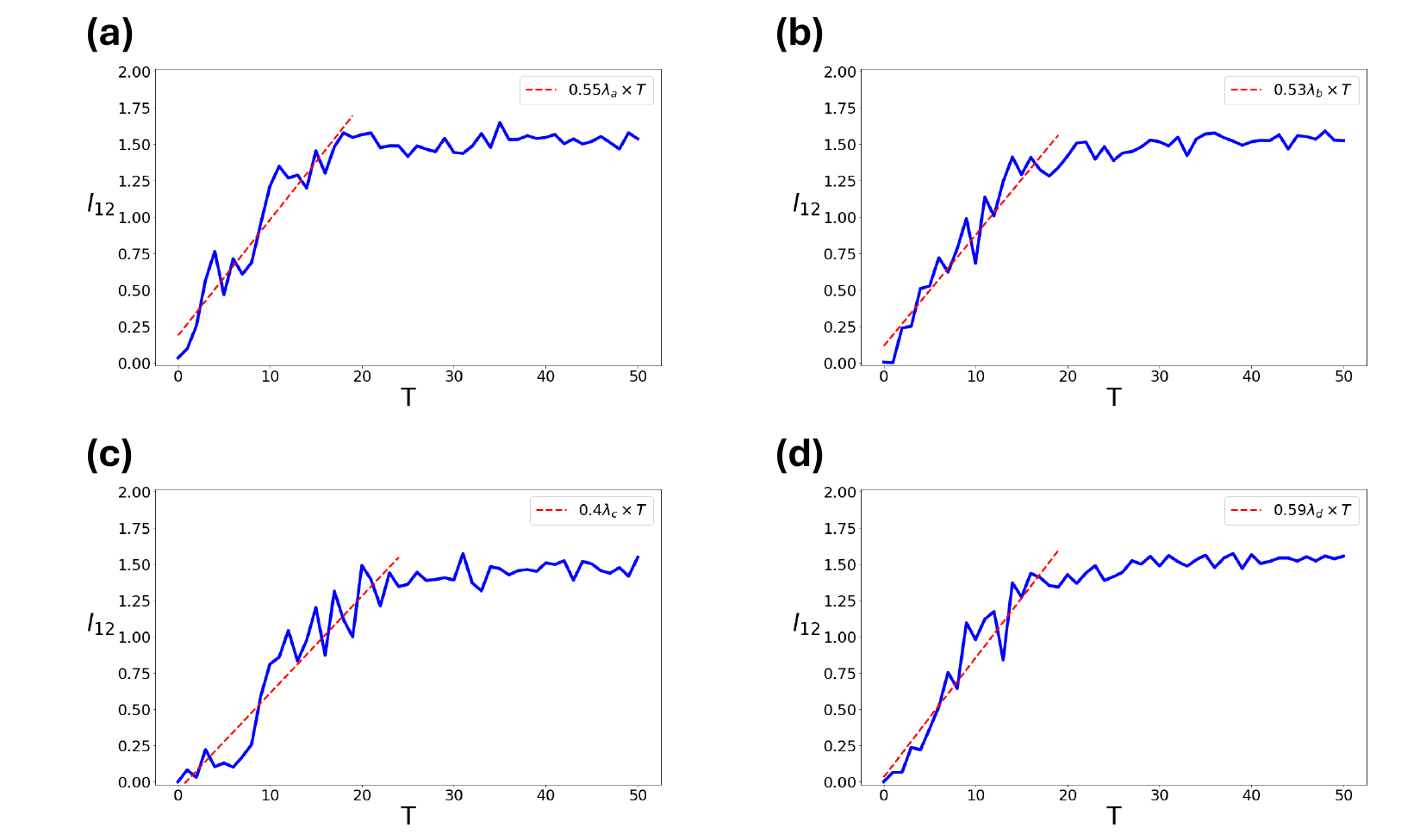}
\caption{\textbf{Mutual information growth and Lyuapunov exponents for a mixed regular-chaotic phase space.} A comparison of the growth rate of $I_{12}$ with the corresponding Lyuapunov exponents at $\kappa=2.5$ for four different chaotic initial conditions: (a) $(\theta_0=3 \pi/4, \phi_0=3 \pi/4)$, $\lambda_a = 0.145$; (b) $(\theta_0=1.0, \phi_0=\pi/10)$, $\lambda_b = 0.143$; (c) $(\theta_0=\pi/5, \phi_0=\pi/10)$, $\lambda_c = 0.167$; (d) $(\theta_0=\pi/4, \phi_0=\pi/3)$, $\lambda_d = 0.139$. For all these scenarios, $j=100$, and $1000$ samples are drawn from the initial distribution.} 
\label{miwithtime2.5}
\end{figure}

In Fig.~\ref{miwithsize}(b), we have plotted the dynamics of $I_{12}$ for chaotic classical dynamics. Clearly, $I_{12}$ grows almost linearly once the initial transient subsides, during which it does not grow at all, for larger $j$. Moreover, plot (d) shows that the equilibration time $T_{eq}$ for $I_{12}$, just like quantum entropy, increases with $j$ much more slowly as $O (\ln j)$ compared to the regular case. 

The uniform growth rate for quantum entropy under chaotic classical dynamics is known to be proportional to the positive Lyuapunov exponent\footnote{More generally, the growth rate is proportional to the sum of the positive Lyuapunov exponents, but for the kicked top, there is just one such exponent.} of the corresponding classical dynamics\cite{ktm-eePRA}. To check whether this holds for classical mutual information $I_{12}$ as well, we have compared the growth rates of $I_{12}$ with the corresponding Lyuapunov exponents for four randomly chosen initial orientations in a fully chaotic phase space at $\kappa=6.0$. To estimate the Lyuapunov exponents, we used the standard algorithm of Benettin et al\cite{ktm-eePRA, benettin1976, benettin1980a, benettin1980b} (see supporting information \ref{lexponents}.) At $\kappa=6.0$, we expect the Lyuapunov exponent to be nearly uniform throughout the phase space. For all these cases, we find that the growth rate is $\sim 0.5 \times \lambda$, where $\lambda$ is the positive Lyuapunov exponent corresponding to the point $(\theta_0,\phi_0)$.

Similarly, we have also compared the growth rates of $I_{12}$ with the corresponding Lyuapunov exponents for four randomly chosen chaotic initial conditions in a mixed regular-chaotic phase space at $\kappa=2.5$ in Fig.~\ref{miwithtime2.5}. We again find that the growth rate of $I_{12}$ is on the order of $0.5 \times \lambda$; however, the deviations from this value seem to be larger for the mixed regular-chaotic phase space than for the fully chaotic case in Fig.~\ref{miwithtime6.0}.

\begin{figure}[t]
\centering
\includegraphics[width=\linewidth]{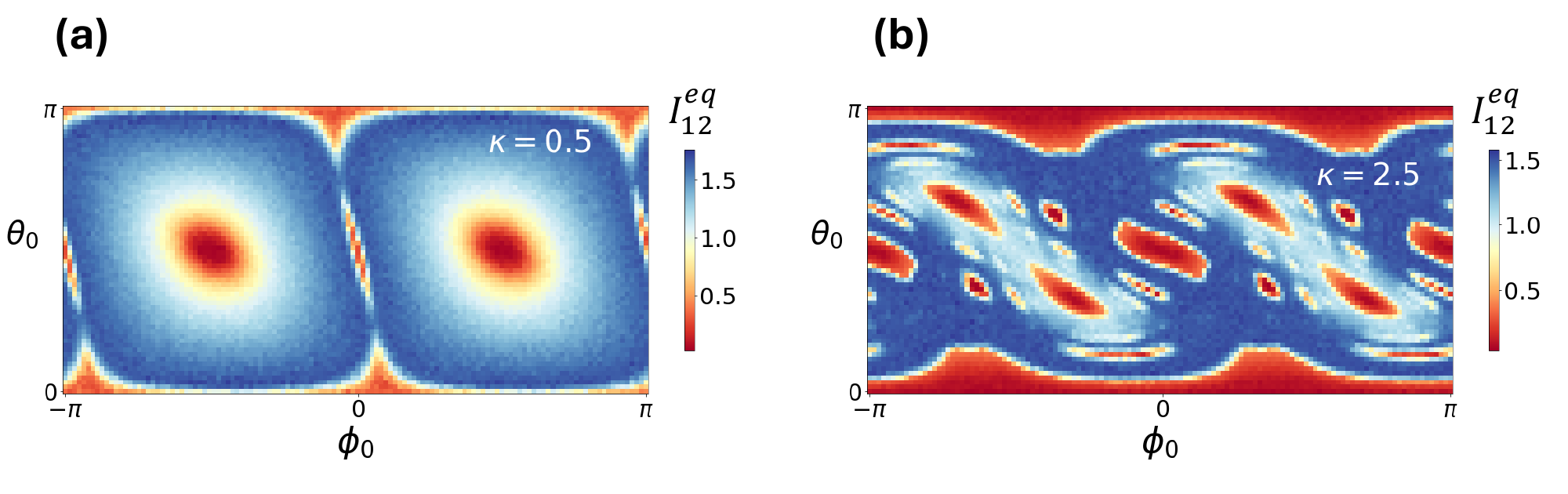}
\caption{\textbf{Equilibrium mutual information.} Equilibrium value of mutual information $I^{eq}_{12}$ is estimated as a function of $(\theta_0,\phi_0)$ for: (a) $\kappa=0.5$ and (b) $\kappa=2.5$. For each $(\theta_0,\phi_0)$, $200$ trajectories are sampled to compute the statistics and $j=100$ is used. To obtain $I^{eq}_{12}$, $I_{12}$ is averaged between $400 \leq T \leq 500$ for both cases. The plots are remarkably similar to equilibrium entropy $S_{eq}$ in Figs.~\ref{linearentropy} and \ref{thermlimit}.} 
\label{miportrait}
\end{figure}

Finally, in Fig.~\ref{miportrait}, we have plotted the equilibrium value of $I_{12}$ as a function of the initial orientation $(\theta_0,\phi_0)$ for regular ($\kappa=0.5$) and chaotic ($\kappa=2.5$) cases, respectively. The equilibrium value $I^{eq}_{12}$ is estimated by averaging $I_{12}$ in the time range $400 \leq T \leq 500$ for both cases. The plots obtained look remarkably similar to the plots of equilibrium entropy $S_{eq}$ in Figs.~\ref{linearentropy} and \ref{thermlimit}. Like quantum entropy, classical mutual information as a function of the initial orientation is also able to reflect the structure of the classical phase space. So, the results in Figs.~\ref{miwithsize}, \ref{miwithtime6.0}, \ref{miwithtime2.5} and \ref{miportrait} show that the signatures of chaos associated with entanglement have clear analogues in the statistical interpretation of classical mechanics.

\section{Summary and Outlook}
\label{summary}

In this paper, we have demonstrated that the signatures of chaos displayed by biparitite entanglement can also be observed through a classical statistical measure. Our measure is based on the mutual information between the marginal phase space densities of individual subsystems. We have evolved this quantity dynamically using the kicked top Hamiltonian. Our results can be summarized as follows: (i) mutual information $I_{12}$ grows logarithmically for regular classical dynamics, whereas, the growth is linear and the rate of growth is proportional to the Lyuapunov exponent for chaotic dynamics; (ii) the equilibration time $T_{eq}$ grows with system size $j=N/2$ as $O(\sqrt{j})$ for regular dynamics and $O(\ln j)$ for chaotic dynamics; (iii) the equilibrium mutual information $I^{eq}_{12}$, estimated by averaging $I_{12}$, is larger for initial conditions that produce chaotic trajectories than those that lead to regular motion for a mixed regular-chaotic phase space. All of these are well-known signatures of chaos in bipartite quantum entanglement\cite{bsanders, ghose, chaudhury, google, ktm-eePRA}.

Although this study has been limited to calculating mutual information for a specific bipartition, an important direction for future research would be to consider the signatures of chaos for more general scenarios, such as equal-sized partitions. Another important future goal is to extend this classical analogy to multipartite measures of entanglement such as the quantum Fisher information that are also known to exhibit signatures of chaos\cite{ktm-eePRA}. Moreover, it would also be critical to consider more non-trivial quantum states beyond the simple spin-coherent states to determine the extent to which quantum behavior can be recovered using classical mechanics. Another interesting exploration could be to compare the dynamics of mutual information with other measures of classical nonseparability\cite{sarkar1998, lakshminarayan2001-1, furuya2004, giulio, eberly}.

Finally, these results might also have implications for the foundations of classical and quantum mechanics. In recent years, Gisin et al. have advanced an alternative interpretation of classical mechanics as an attempt to bridge the conceptual gap between classical and quantum physics \cite{gisin2017,gisin2019,gisin2020-1,gisin2020-2,gisin2021-1,gisin2021-2,gisin2024}. Their basic claim is that the orthodox understanding of classical mechanics takes for granted an assumption that they have called the \textit{principle of infinite precision}; that physical quantities can be specified to an infinite number of digits. Once this assumption is relaxed, they have argued, many features exclusively attributed to quantum physics such as the fundamental role of measurement and the nonseparability of states appear analogously in classical physics too \cite{gisin2024}.

We have argued in this paper that classical nonseparability certainly reveals new connections between classical and quantum realms. Gisin et al. go a step further and allow the possibility for classical nonseparability to be a physically real phenomenon. However, the question will remain unresolved until experimental investigations are carried out. One possible route could be to monitor the motion of charged particles in classical and quantum wells \cite{ktm-exp}. Charges moving between parallel planar potential barriers under a magnetic field tilted with respect to the barriers exhibit chaotic dynamics. The emergence of chaos in this system is described by the kicked top map in certain regimes. In the classical version of the system, chaos is accompanied with a large energy transfer between the longitudinal and the cyclotron motion of the charges; however, this energy exchange is suppressed in the quantum limit \cite{ktm-exp}. Further analysis will be needed to investigate the possible experimental signatures of classical nonseparability in this system.

\section*{Acknowledgement}
B.K. would like to acknowledge the contribution of the workshops led by Prof.~Basit Bilal Koshul under \textit{Acacia Education Foundation} in shaping the conceptual outlook of this work. Both authors would like to thank Prof.~Nicolas Gisin, Prof.~Nima Lashkari and Dr.~Mudassir Moosa for many useful suggestions and discussions. We also thank the referees for their constructive comments enabling us to improve our manuscript. The work was supported by the Office of Science through the Quantum Science Center (QSC), a National Quantum Information Science Research Center, and the U.S. Department of Energy (DOE) (Office of Basic Energy Sciences), under Award No. DE-SC0019215.

\section*{Data Availability} 
All data are provided and the computer codes used to generate this study are available\cite{data_availability}.

\pagebreak

\bibliography{references}

\pagebreak

\setcounter{section}{0}
\renewcommand{\thesection}{S-\Roman{section}}

\begin{suppinfo}

\section{Update of Angular Momentum}
\label{recursion}
To compute $\textbf{J}'_1 = U^\dagger \textbf{J}_1 U$, we first note that $U_{z^2}^\dagger \textbf{J}_1 U_{z^2}$ is as follows\cite{ktm-haake}
\begin{align}
    U_{z^2}^\dagger J_{1x} U_{z^2} &= \frac{1}{2} (J_{1x} + \diota J_{1y}) \: \text{e}^{ \diota \frac{\kappa}{j} (J_{1z} + \frac{1}{2})} + \text{h.c.} \nonumber \\
    U_{z^2}^\dagger J_{1y} U_{z^2} &= \frac{1}{2 \diota} (J_{1x} + \diota J_{1y}) \: \text{e}^{ \diota \frac{\kappa}{j} (J_{1z} + \frac{1}{2})} + \text{h.c.} \label{cpstep1} \\
    U_{z^2}^\dagger J_{1z} U_{z^2} &=  J_{1z} \nonumber.
\end{align}
On the other hand, $U_{12}^\dagger \textbf{J}_1 U_{12}$ simply rotates $\textbf{J}_1$ in the following way
\begin{align}
    U_{12}^\dagger J_{1x} U_{12} &= J_{1x} \cos \Big(\kappa \frac{J_{2z}}{j} \Big) - J_{1y} \sin \Big(\kappa \frac{J_{2z}}{j} \Big) \nonumber \\
    U_{12}^\dagger J_{1y} U_{12} &= J_{1x} \sin \Big(\kappa \frac{J_{2z}}{j} \Big) + J_{1y} \cos \Big(\kappa \frac{J_{2z}}{j} \Big) \label{cpstep2} \\
    U_{12}^\dagger J_{1z} U_{12} &= J_{1z} \nonumber.
\end{align}
Combining \eqref{cpstep1} and \eqref{cpstep2}, we get
\begin{align}
    U_{12}^\dagger U_{z^2}^\dagger J_{1x} U_{z^2} U_{12} &= \frac{1}{2} (J_{1x} + \diota J_{1y}) \: \text{e}^{ \diota \frac{\kappa}{j} (J_{1z} + J_{2z} + \frac{1}{2})} + \text{h.c.} \nonumber \\
    U_{12}^\dagger U_{z^2}^\dagger J_{1y} U_{z^2} U_{12} &= \frac{1}{2 \diota} (J_{1x} + \diota J_{1y}) \: \text{e}^{ \diota \frac{\kappa}{j} (J_{1z} + J_{2z} + \frac{1}{2})} + \text{h.c.} \\
    U_{12}^\dagger U_{z^2}^\dagger J_{1z} U_{z^2} U_{12} &=  J_{1z} \nonumber.
\end{align}
Finally, performing the rotation around the y-axis gives us the following answer for $\textbf{J}'_1 = U^\dagger \textbf{J}_1 U = U_y^\dagger U_{12}^\dagger U_{z^2}^\dagger \textbf{J}_1 U_{z^2} U_{12} U_{y}$,
\begin{align}
    J'_{1x} &= \frac{1}{2} (J_{1z} + \diota J_{1y}) \: \text{e}^{- \diota \frac{\kappa}{j} (J_{1x} + J_{2x} + \frac{1}{2})} + \text{h.c.} \nonumber \\
    J'_{1x} &= \frac{1}{2 \diota} (J_{1z} + \diota J_{1y}) \: \text{e}^{- \diota \frac{\kappa}{j} (J_{1x} + J_{2x} + \frac{1}{2})} + \text{h.c.} \\
    J'_{1x} &=  - J_{1x} \nonumber.
\end{align}

\section{Oscillatory Growth of Mutual Information}
\label{oscilaltionsmi}
In this section, we analyze the oscillatory growth of mutual information $I_{12}$ for regular initial conditions as shown in Fig.~\ref{miwithsize}(a). In Fig.~\ref{oscillations}, we have replotted the mutual information growth for the initial conditions used in Fig.~\ref{miwithsize}(a) by increasing the number of nearest neighbors in the data used to estimate mutual information from $k=3$ to $k=10$\cite{mi_alg}. Moreover, we also increased the number of trajectories that were sampled from $500$ to $1000$. However, the oscillations still persist, implying that these oscillatory features are not results of sampling and estimation inaccuracies. We observe similar oscillatory behavior for a different initial condition in Fig.~\ref{oscillations}(b). Thus, this oscillatory growth seems to be a characteristic feature for regular initial conditions.

In Figs.~\ref{oscillations}(c) and (d), we have plotted the growth of linear entropy in the thermodynamic limit $j \to \infty$ for two different initial conditions using $S = \ev{(\Delta X)^2}/2$ where $\ev{(\Delta X)^2} = (\ev{\textbf{J}^2} - \ev{\bf{J}}^2) / j^2$ is estimated classically. For both calculations, the initial points were sampled uniformly from a region of angular spread $\sin \theta_0 \Delta \theta \Delta \phi=1/j$ with $j=1000$ centered on the respective initial points. A more pronounced oscillatory growth is observed in these figures compared to Fig.~\ref{linearentropy}(a) which suggests that this oscillatory growth is not an exclusively classical feature but is observed in quantum entropy as well for large enough systems.

\begin{figure}
\centering
\includegraphics[width=\linewidth]{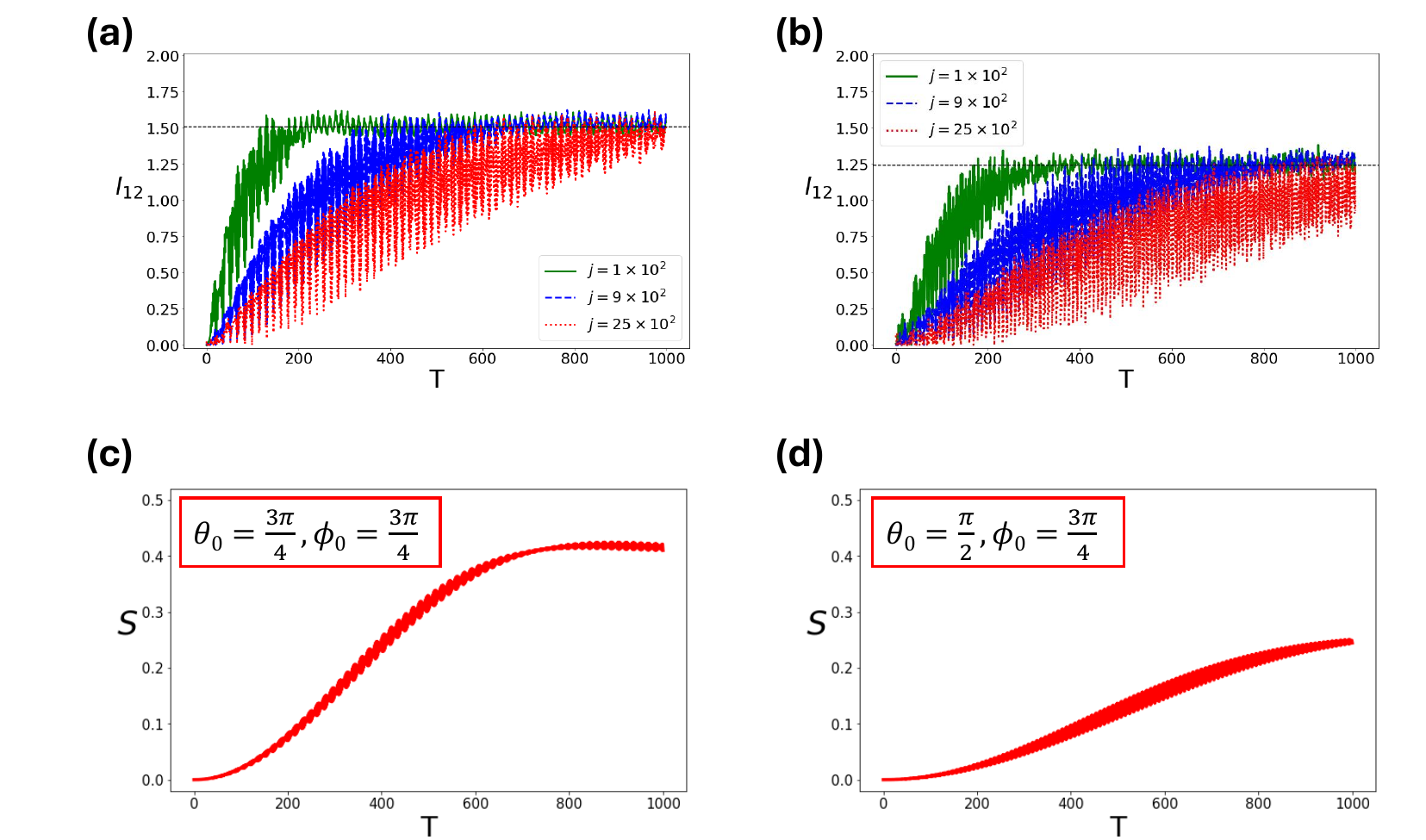}
\caption{\textbf{Oscillatory growth of mutual information.} Oscillatory growth of mutual information $I_{12}$ for regular initial conditions. (a) $I_{12}$ is replotted for the initial conditions in Fig.~\ref{miwithsize}(a) i.e. $(\theta_0=3 \pi/4,\phi_0=3 \pi/4)$, with $k=10$ neighboring data points used for entropy estimation compared to $k=3$ used in Fig.~\ref{miwithsize}(a). Moreover, $1000$ samples were drawn and evolved from the initial distribution compared to only $500$ for Fig.~\ref{miwithsize}(a). The robustness of these oscillations against improvements in estimation implies that they cannot be a result of sampling and estimation inaccuracies. (b) The growth of $I_{12}$ for $(\theta_0=\pi/2,\phi_0=3 \pi/4)$ again with $\kappa=0.5$. For this case, we used $k=3$ and sampled $500$ points. The oscillatory growth seems to be a characteristic feature for all regular initial conditions. (c) \& (d) Estimate of linear entropy $S$ using $S = \ev{(\Delta X)^2}/2$ where $\ev{(\Delta X)^2} = (\ev{\textbf{J}^2} - \ev{\bf{J}}^2) / j^2$ is computed classically with initial points uniformly sampled from a region of angular spread $\sin \theta_0 \Delta \theta \Delta \phi=1/j$ with $j=1000$. For both plots, we sampled $200$ initial points. We observe more pronounced oscillations for $S$ here than in Fig.~\ref{linearentropy}(a).} 
\label{oscillations}
\end{figure}

\section{Calculation of Lyuapunov Exponents}
\label{lexponents}
In this section, we apply the procedure of Benettin et al.\cite{ktm-eePRA,benettin1976,benettin1980a,benettin1980b} to estimate the Lyuapunov exponents for the kicked top map \eqref{crecur}. Suppose our initial point is $\textbf{X}_0 = (\sin \theta_0 \cos \phi_0, \sin \theta_0 \sin \phi_0, \cos \theta_0)$. First, we pick two independent tangent vectors $(\textbf{W}^{(1)}_0,\textbf{W}^{(2)}_0)$ at the point $\textbf{X}_0$ on the unit sphere. These vectors can be chosen at random. For our calculations, we choose
\begin{align}
    \textbf{W}^{(1)}_0 &= \begin{bmatrix}
           \cos \theta_0 \cos \phi_0 \\
           \cos \theta_0 \sin \phi_0 \\
           - \sin \theta_0 
         \end{bmatrix}; \: \: \: \: \: \: 
    \textbf{W}^{(2)}_0 = \begin{bmatrix}
           \sin \phi_0 \\
           -\cos \phi_0\\
           0 
         \end{bmatrix}.
\end{align}
$\textbf{X}_i$ is updated through $\textbf{X}_{i+1} = (F_X [\textbf{X}_{i}], F_Y [\textbf{X}_{i}], F_Z [\textbf{X}_{i}])$ where $F_X$, $F_Y$ and $F_Z$ are given in eqs.~\eqref{crecur}. The tangent vectors are updated using the map $\textbf{W}^{(1)}_{i+1} = \textbf{A} [\textbf{X}_i] \: \textbf{W}^{(1)}_{i}$ where
\begin{align}
    \textbf{A} [\textbf{X}_i] &= \begin{bmatrix}
        \partial_{X_i} F_X [\textbf{X}_i] & \partial_{Y_i} F_X [\textbf{X}_i] & \partial_{Z_i} F_X [\textbf{X}_i] \\ 
        \partial_{X_i} F_Y [\textbf{X}_i] & \partial_{Y_i} F_Y [\textbf{X}_i] & \partial_{Z_i} F_Y [\textbf{X}_i] \\ 
        \partial_{X_i} F_Z [\textbf{X}_i] & \partial_{Y_i} F_Z [\textbf{X}_i] & \partial_{Z_i} F_Z [\textbf{X}_i]
    \end{bmatrix}.
\end{align}
The procedure to obtain the Lyuapunov exponent is as follows\cite{ktm-eePRA}:
\begin{enumerate}
    \item Evolve the tangent vectors $(\textbf{W}^{(1)}_{(i-1)s},\textbf{W}^{(2)}_{(i-1)s})$ for $s$ time steps to $(\textbf{W}^{(1)}_{is},\textbf{W}^{(2)}_{is})$.
    
    \item Apply the Gram-Schmidt procedure:
    \begin{align}
    \alpha_i &= |\textbf{W}^{(1)}_{is}|, \: \: \: \: \: \textbf{V}^{(1)} = \textbf{W}^{(1)}_{is} / \alpha_i; \\
    \beta_i &= |\textbf{W}^{(2)}_{is} - (\textbf{V}^{(1)} \cdot \textbf{W}^{(2)}_{is}) \textbf{V}^{(1)}|, \: \: \: \: \: 
    \textbf{V}^{(2)} = \frac{1}{\beta_i} [\textbf{W}^{(2)}_{is} - (\textbf{V}^{(1)} \cdot \textbf{W}^{(2)}_{is}) \textbf{V}^{(1)}].
    \end{align}

    \item Reinitialize $\textbf{W}^{(1)}_{is} = \textbf{V}^{(1)}$ and $\textbf{W}^{(2)}_{is} = \textbf{V}^{(2)}$.
\end{enumerate}
Then, for large $n$, an estimate of the postive Lyuapunov exponent $\lambda$ is obtained through
\begin{equation}
    \lambda^{(n,s)} = \frac{1}{ns} \sum_{i=1}^{n} \ln \alpha_i.
\end{equation}
This expression converges to $\lambda$ in the limit $n \to \infty$. Figs.~\ref{lexp6.0} and \ref{lexp2.5} show the convergence of Lyuapunov exponents for the corresponding scenarios in Figs.~\ref{miwithtime6.0} and \ref{miwithtime2.5} respectively.

\begin{figure}[t]
\centering
\includegraphics[width=\linewidth]{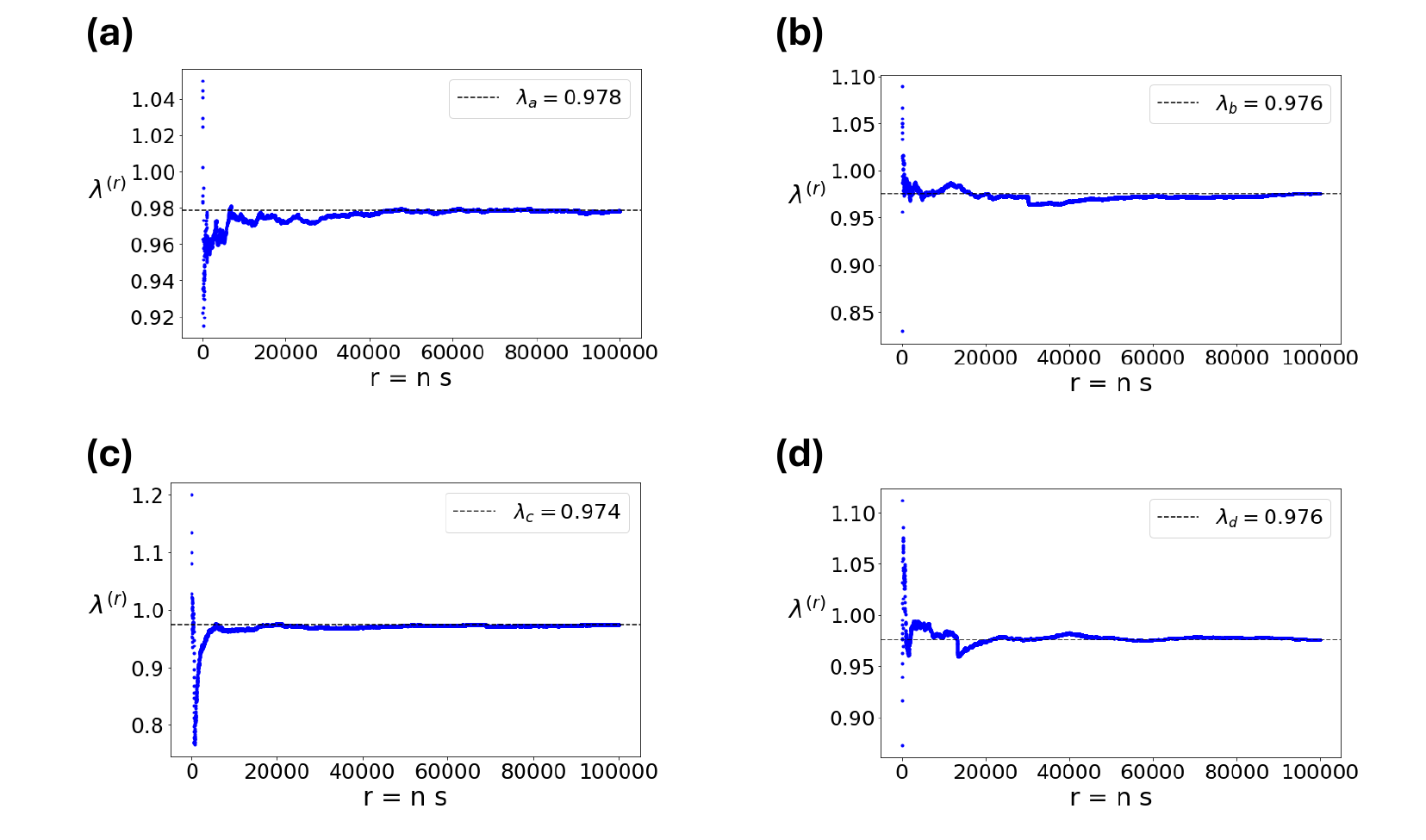}
\caption{\textbf{Convergence of Lyuapunov exponents at $\kappa=6.0$.} Convergence of Lyuapunov exponents for the four scenarios of Fig.~\ref{miwithtime6.0} with $s=10$: (a) $(\theta_0=3 \pi/4, \phi_0=3 \pi/4)$, $\lambda_a = 0.978$; (b) $(\theta_0=\pi/3, \phi_0=2 \pi/3)$, $\lambda_b = 0.976$; (c) $(\theta_0=1.0, \phi_0=\pi/10)$, $\lambda_c = 0.974$; (d) $(\theta_0=\pi/4, \phi_0=\pi/3)$, $\lambda_d = 0.976$.} 
\label{lexp6.0}
\end{figure}

\begin{figure}[t]
\centering
\includegraphics[width=\linewidth]{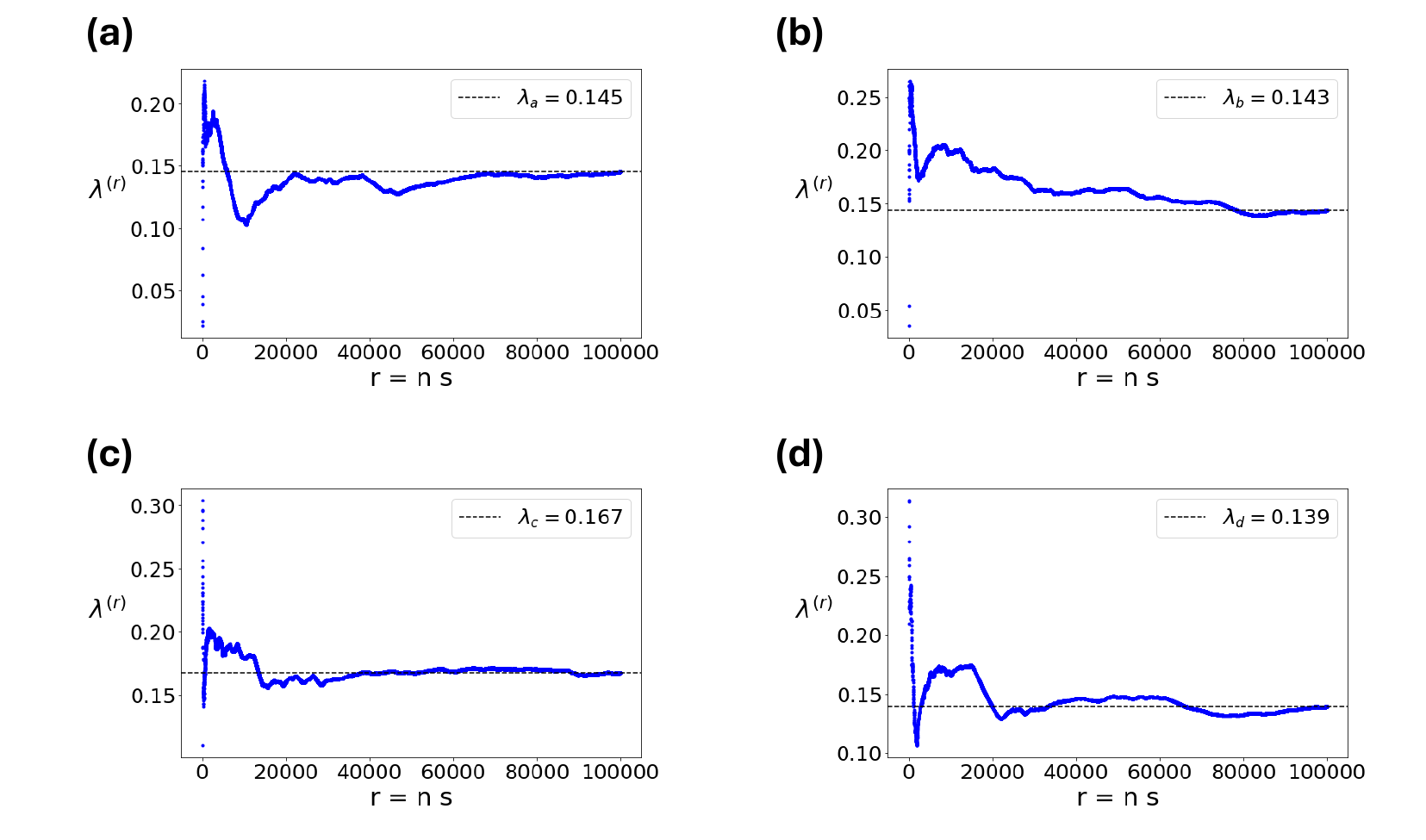}
\caption{\textbf{Convergence of Lyuapunov exponents at $\kappa=2.5$.} Convergence of Lyuapunov exponents for the four scenarios of Fig.~\ref{miwithtime2.5} with $s=5$: (a) $(\theta_0=3 \pi/4, \phi_0=3 \pi/4)$, $\lambda_a = 0.145$; (b) $(\theta_0=1.0, \phi_0=\pi/10)$, $\lambda_b = 0.143$; (c) $(\theta_0=\pi/5, \phi_0=\pi/10)$, $\lambda_c = 0.167$; (d) $(\theta_0=\pi/4, \phi_0=\pi/3)$, $\lambda_d = 0.139$.} 
\label{lexp2.5}
\end{figure}

\end{suppinfo}

\end{document}